\documentstyle[12pt, aasms4, epsfig]{article}
\catcode`@=11
\def\eqalign#1{\null\,\vcenter{\openup\jot\m@th
  \ialign{\strut\hfil$\displaystyle{##}$&$\displaystyle{{}##}$\hfil
      \crcr#1\crcr}}\,}
\def\eqalignleft#1{\null\,\vcenter{\openup\jot\m@th
  \ialign{\strut$\displaystyle{##}$\hfil&$\displaystyle{{}##}$\hfil
      \crcr#1\crcr}}\,}
\catcode`@=12

\def\lax    {\ifmmode{_<\atop^{\sim}}\else{${_<\atop^{\sim}}$}\fi}
\def\gax    {\ifmmode{_>\atop^{\sim}}\else{${_>\atop^{\sim}}$}\fi}
\def\kms    {\ifmmode{{\rm ~km~s}^{-1}}\else{~km~s$^{-1}$}\fi}
\def\kmp    {~km~s$^{-1}$~pc$^{-1}$}
\def\lo     {~${\rm L}_{\odot}$}
\def\mo     {~${\rm M}_{\odot}$}
\def\moyr   {\hbox{~$M_{\odot}\,{\rm yr}^{-1}$}}
\def\etal   {{\sl et~al.~}}
\def\eg{{\it e.\thinspace g.~}}
\def\ie{{\it i.\thinspace e.~}}
\def\bk{\lower 6pt\hbox{${\buildrel k\over \sim}$}}
\def\bv{\lower 6pt\hbox{${\buildrel v\over \sim}$}}
 
\def\muas   {~$\mu${\rm as}}
\def\masyr  {~mas~yr$^{-1}$}
\def\htwoo  {~${\rm H_2O}$}
\def\vbot   {$V_\bot$}
\def\blankline  {\vskip10truept}
\def\sal {\vskip 12 truept}

\begin{document}
\doublespace
\textwidth 6.5truein
\textheight 9.25truein
\topmargin -1cm
\hoffset = 0.5truein
\title{\bf 
FOUR COMETARY BELTS\\ ASSOCIATED WITH THE ORBITS OF
GIANT PLANETS:\\ A New View of the Outer Solar System's Structure
Emerges From Numerical Simulations 
}
\author{\bf Leonid M. Ozernoy\altaffilmark{1}}  
\affil{5C3, Computational Sciences  Institute and Department of Physics 
\& Astronomy,\\ George Mason U., Fairfax, VA 22030-4444; also Laboratory for 
Astronomy and Solar\\ Physics, NASA/Goddard Space Flight Center, Greenbelt, 
MD 20771} 
\altaffiltext{1}{Corresponding author. Fax: $+1$-301-286-1617; e-mail: 
ozernoy@science.gmu.edu, ozernoy@stars.gsfc.nasa.gov}
\author{\bf Nick N. Gorkavyi\altaffilmark{2}}
\affil{
Laboratory for Astronomy and Solar Physics, NASA/Goddard Space Flight 
Center\\ Greenbelt, MD 20771; also 
Simeiz Department, Crimean Astrophysical Observatory, Simeiz 334242, Ukraine
}
\altaffiltext{2}{NRC/NAS Senior Research Associate; e-mail: 
gorkavyi@stars.gsfc.nasa.gov}
\author{\bf Tanya Taidakova\altaffilmark{3}}
\affil{
Computational Consulting Service, College Park, MD 20740
}
\altaffiltext{3}{e-mail: simeiz@aol.com}
\bigskip

\begin{abstract}
Using numerical simulations, we examine the 
structure of a cometary population near a massive planet, such as
a giant planet of the Solar system, starting with one-planet approximation
(the Sun plus one planet). By studying the distributions of comets 
in semimajor axis, eccentricity, pericenter, and apocenter distances, 
we have revealed several interesting features in these distributions. 
The most remarkable ones include (i) spatial accumulation of comets
near the planetary orbit (which we call the {\it `cometary belt'}) 
and (ii) avoidance of resonant orbits by comets.
Then we abandon one-planet approximation and examine as to how a cometary 
belt is modified when the influence of all four giant planets is taken into
consideration. To this end, we simulate a stationary distribution of 
comets, which results from the gravitational scattering of the Kuiper 
belt objects on the four giant planets and accounts for the effects 
of mean motion resonances. In our simulations, we deal with the stationary 
distributions computed, at different initial conditions, as 36 runs for 
the dynamical evolution of comets, which start from the Kuiper belt and 
are typically traced until the comets are ejected from the Solar system. 
Accounting for the influence of four giant planets makes the
cometary belts overlapping, but nevertheless keeping almost all their basic 
features found in one-planet approximation. In particular, the belts 
maintain the gaps in the $(a,e)$- and $(a,i)$-space similar to
the Kirkwood gaps in the main asteroid belt. We conclude that the large-scale
structure of the Solar system is featured by the four cometary belts
expected to contain 20-30 millions of scattered comets, and only 
a tiny fraction of them is currently visible as Jupiter-, Saturn-, etc. 
family comets. 
\end{abstract}

\newpage
\section*{1. Introduction}
    
The outer Solar system beyond the four giant planets includes the Kuiper belt
and the Oort cloud, which contain a raw material left since the formation
of the system. The Kuiper belt objects 
are thought to be responsible for progressive 
replenishment of the observable cometary populations, and gravitational 
scattering of these objects on the four giant planets could provide their
transport from the trans-Neptunian region all the way inward, down to 
Jupiter (Fern\'andez \& Ip, 1983; Torbett 1989; Levison \& Duncan, 1997; for
a review, see Malhotra, Duncan \&  Levison, 1999 and references therein). 
The present paper, which continues and develops an approach started by 
Ozernoy, Gorkavyi, \& Taidakova 2000 ($\equiv$ OGT), makes the
emphasis on the structure of cometary populations between Neptune and 
Jupiter, both in phase space, i.e. in the space of orbital 
elements \{$a,e,i$\}, and in real space. 
We argue that there are spatial accumulations of comets near the 
orbits of all four giant planets, which we name {\it cometary belts}. 
These populations have the 
dynamical nature, because the comets belonging to the given planet's belt 
are either in a resonance with the host or are gravitationally scattered 
predominantly on this planet.

 Our approach (which has a number of common elements with the 
`particle-in-cell' computational method) is, in brief, as follows: 
Let us consider, for simplicity, a stationary particle distribution
in the frame co-rotating with the planet (Neptune). The locus of the
given  particle's  positions (taken, say, as $6\cdot 10^3$ positions
every $10^6$ yrs, i.e. every $6\cdot 10^3$ Neptune's revolutions about 
the Sun) are recorded and considered as the
positions of {\it many other particles} of  the same origin 
 but {\it at a different  time}. After this particle `dies' (as a result
of infall on a planet/the Sun or ejection from the system by a 
planet-perturber), its recorded positions
sampled over its lifetime form a stationary distribution as if it were
produced by {\bf many} particles. Typically, each run includes  
$0.7\cdot 10^6$ Neptune's revolutions ($10^8$ yrs) to give  
$\sim 0.7\cdot 10^6$ positions of a particle, which is
equivalent, for a stationary distribution, to the same number of particles. 

We integrate the dynamical equations for the motion of a massless particle
in the gravitational field of the Sun and the four giant planets written in the
rotating $\{x,y,z\}$-coordinate system with the $x$-axis directed along the
radius and the $y$-axis in the direction of the orbital motion of Neptune,
while the origin of coordinates is placed at the center of the Sun (thereby the
secular resonances are not considered here). We use an implicit second-order
integrator (Taidakova 1997) appropriately adapted to achieve our goals
are described in detail in OGT (2000); as shown there, the integrator
for a dissipationless system provides the necessary accuracy of
computations on the time scale of $0.5\cdot 10^9$ years. A big advantage
of this integrator is its stability: an error in the energy (the Tisserand
parameter) does not grow as the number of time steps increases
if the value of the step remains the same. The latter
situation is exemplified by a resonant particle -- it
does not approach too close to the planet so that the same time
step can be taken. In contrast to resonant particles,
non-resonant ones, in due course of their
gravitational scatterings, approach one or another planet from time to time,
and therefore one has to change the time step near the planet. Obviously, 
whenever the time step diminishes near
the planet, an error in the Tisserand parameter slowly grows together with
an increased number of the smaller time steps. Nevertheless, in our
simulations a fractional error in the Tisserand parameter typically does
not exceed 0.001 during $3\cdot 10^6$ Neptune's revolution, which amounts
0.5 Gyrs (OGT 2000). To increase accuracy of computations,
we use in the present paper a second iteration. While the 1st iteration
yields the gravitational field between points $A$ and $B$ using an
approximative formula based on the  particle parameters at point $A$
(because those at point $B$ are still unknown), the 2nd iteration
enables us to compute the gravitational field between $A$ and $B$ 
using a middle position between them because
the position of $B$ is already given by the 1st iteration.

We commence this paper with examining the characteristics of
cometary belts  in one-planet approximation, i.e. in a 3-body problem:
the Sun, a planet, and a comet (Sec.~2). In Sec.~3, we consider
how these cometary belts are modified when the influence of all four
giant planets is taken into account. Sec.~4 contains discussion and 
conclusions.

\section*{2. One-planet approximation}
    
We consider as an example a giant planet of the Saturnian mass placed on
a circular orbit of radius $R_{pl}$ having a zero inclination. It is assumed
that there is an outer source of comets, which injects them into the  
Saturn's strong scattering zone. The boundaries of the latter are 
close to the region, where heliocentric orbits of comets cross the planet's
orbit: this region, which we call hereinafter the {\it crossing zone}, is 
defined as $a(1-e)\leq R_{pl}\leq a(1+e)$. An outer source of comets could 
be the Kuiper belt, a scattering zone of an outer 
planet, or (at a much earlier stage) the disk of planetesimals.
     
The dynamical evolution of comets typically includes multiple
gravitational scatterings of comets on the planet with eventual ejection
from the system with a hyperbolic velocity. (If neighboring planets
are included, some comets may enter the 
scattering zones of those planets).  On rare occasions, there are
impacts of comets with the planet. 
     
We have computed the dynamical evolution of test comets by making
a record of their orbital parameters each revolution of the planet. 
Assuming that the inflow of comets into the planet's scattering zone 
does not change in time, we can interpret the sample of orbital
parameters of test bodies as representing  a stationary distribution 
of a large number of comets.

We simulated 26 distributions of cometary orbits totalling
$0.37\times 10^6$ positions in the form of $(a,e,i)$-points.
The initial conditions for integration of cometary orbits were 
taken out of resonances: $a_0=1.1~ a_{planet}=10.5$ AU (for $e_0$ and
$i_0$, see Table 1, where details of this
and other computational runs are given as well).

Earlier (OGT) we computed the stationary distribution of test
comets in the space of orbital coordinates $(a,e)$ and $(a,i)$. 
As distinct from OGT (1999), where each comet was represented as a point
on the phase plane [$(a,e)$ or $(a,i)$], in the present paper we wish 
to compute the distribution of the `surface density'
(more accurately, the 2D-density) of comets on the
phase plane. To this end, we make a record of
coordinates of test comets each 10 revolutions of the planet.
Then we sort out the computed $0.37\times 10^{6}$  cometary coordinates 
into two 2D data files: a $500\times 100$ array in the $(a,e)$-plane 
($a <2.5 R_{pl}$) and a $500\times 180$ array in the $(a,i)$-space). 
The following bins are used: $\Delta a = 0.005 R_{pl}, ~\Delta e =0.01, 
~\Delta i=0.5^\circ$. 
Fig.~1a,b show, in the $(a,e)$ and $(a,i)$-spaces, the surface density of 
comets, which are gravitationally scattered on a planet of the Saturnian mass.
The following features are worth mentioning: 
\begin{description}
\item {\it (i)} 
in the $(a,e)$-space, the comets are stretched along the boundaries 
 of the planet's strong scattering zone;

\item {\it (ii)} 
 resonant gaps at the resonances 2:3, 1:1, 3:2, 2:1, 3:1, etc., 
 are well pronounced;

\item {\it (iii)} 
 outside the scattering zone, the dynamical evolution of test bodies occurs 
 slowly (in a diffusive way) resulting in clusterings, which could be named
{\it diffusive accumulations}. As can be seen in Fig.~1a, these accumulations 
are separated from the the right boundary $a(1-e)=R_{pl}$ of the crossing
zone by a noticeable `trough' of a decreased surface density of comets.
\end{description}
 
Fig.~2 a,b,c,d show distributions of comets in semimajor axis, pericenter
distance, apocenter  distance, and heliocentric distance, respectively.
The vertical coordinate is a measure of a number of comets within the bin of 
0.01~$R_{pl}$. Dash line delineates the region occupied by comets with
distances of pericenter $<0.5R_{pl}$, i.e. those comets whose chances to be
discovered are the best. Such objects could be called `visible comets' (which
is correct for Jupiter, although for more distant planets the visibility
condition should be more stringent).

 Fig. 2a reveals a rich resonant structure of the cometary belt. Arrows 
 show positions of particular resonances. A detailed analysis (OGT)
 indicates that the smaller the mass of a planet (in the range
 $M_{Uranus} < M < M_{Jupiter}$), the more rich is its resonant structure.
This is caused by the fact that, for large planetary masses, different 
resonances are partly overlapping.

 Fig. 2b demonstrates that the pericenter distances of the scattered comets 
 are located close to the orbit of the planet, slightly exceeding it.
 This is determined by dynamics of comets 
 and is an important feature of the cometary belt.

The distribution of comets in apocenter distance (Fig. 2c)
 demonstrates an appreciable
 concentration of comets to the planet's orbit. If we only select
 comets with {\it small} distances of pericenter (as mentioned above,
it is the condition to have these comets visible),
 it turns out that the apocenter  distances
 of those comets are indeed rather close to the orbit of the planet (see
dashed  curve in Fig.~2c). As is known, it is this circumstance which
   has been used to define the particular `cometary family'.

Fig.~2d (the distribution in heliocentric distance) demonstrates that
 a strong concentration of comets toward the planetary orbit exists 
not only in phase space, but in usual space as well.  
 The cometary belt has a pronounced  maximum near the  host planet's orbit.
Interestingly, the simulated distributions of surface density of both visible 
and all
comets have two maxima. As for the curve of visible comets, this is explained
by the fact that the probability to find a comet at various distances from
the Sun has two maxima -- at the pericenter and apocenter, and since the
visible comets have quite similar orbital parameters those two maxima are 
clearly revealed. As for the simulated comets belonging to the same cometary
belt, the right (outer) peak on the surface density curve appears because
the pericenter distances of all outer (relative to the planet) comets
are rather close to each other, whereas their apocenter distances differ
substantially. Meanwhile the left (inner) peak 
on the surface density curve appears because
the apocenter distances of all inner (relative to the planet) comets
are rather close to each other, whereas their pericenter distances are very
different.
Our simulations indicate that the regions of the largest surface density
of comets are located slightly outside the boundaries of the crossing zone 
(see Fig.~1a). Therefore pericenters of the outermost
comets do not coincide with apocenters of the innermost comets. As a result,
this leads to two maxima in the surface density of comets near the planet's
orbit.

Distributions shown in Figs.~2a to 2d make it quite convincing that {\it
there is a spatial accumulation of comets near the planetary orbit}. We name
such an accumulation the {\it cometary belt}. This population has the 
dynamical nature, because the comets belonging to the given planet's belt 
are either in a resonance with the host or are gravitationally scattered 
predominantly on this planet. 

Summarizing similarities and dissimilarities between 
the cometary belt and cometary family, we can conclude that the
distribution in apocenter distance is the only one which looks alike for 
both, whereas all other distributions are different due to observational
selection to which the cometary family is highly sensitive.

The above material concerns the cometary belt around a planet of the 
Saturnian mass. We have computed the surface density distribution of comets
around a planet of the Jovian, Uranian, and Neptunian mass as well.
The details of our computational runs are given in Table~1.
The data on the surface density of comets for all for giant planets,
in one-planet approximation,
are shown in Fig.~3. As can be seen, the smaller the planet's 
mass, the larger is the surface density contrast.

\section*{3. Four-planet approximation}

It would be important to see which features of cometary belts are kept
invariable when we abandon one-planet approximation and take into account
gravitational fields of all four giant planets. However, to save the
computational time, we continue to neglect eccentricities and inclinations
of the planets (as shown in OGT, accounting for non-zero planetary
eccentricities does not lead to any oversimplifications) and we also neglect
secular resonances.
We have simulated 36 distributions of cometary orbits totalling $25.7\times
10^6$ positions, or $(a,e,i)$-points. The details of our computational runs
are summarized in Table~1.
The initial conditions for orbit integrations are taken in the  
Kuiper belt: we consider  those objects 
whose orbits intersect the Neptune's orbit. They
belong to the resonances 3:2 and 2:1, but the angle `the test body -- the
Sun -- Neptune' was taken in such a way that the test body rapidly leaves
the resonance owing to a close encounter with the planet. Available
numerical computations (Malhotra et al. 1999 and refs. therein) confirm
that the above resonances are temporary and their de-population might
indeed explain the origin of the so called scattered disk objects.
    
Fig.~4a,b show the `surface density' (more accurately, the 2D-density)
of comets in the
$(a,e)$- and $(a,i)$-space governed by gravitational scatterings on 
all four giant planets. 

We use, as we did before in Fig.~1, the logarithmic grey scale, 
with the only difference 
that each shade differs 100-fold from the neighboring one. 
The basic time step used is 0.001 (in units of one Neptune's revolution,
taken to be $2\pi$).
The time step was taken smaller as the test comet approaches the planet.
These improved simulations provide a 4-fold better accuracy compared to
our earlier approach (OGT), where we used a larger basic time step (0.002).
The results of both simulations turn out to be close to each other.

As can be seen from Fig.~4a, the resonant gaps in Neptune's zone as well as
gaps in the resonances 1:1 near Uranus, Saturn and Jupiter are 
 well pronounced (at not too large eccentricities).

Our simulations indicate a progressive, sharp decrease in surface density
of comets between orbits of Neptune and Jupiter (see Fig.~5d).
This decrease is characterized by the transfer functions computed in OGT.
We notice that a substantial decrease in surface density of comets
between orbits of Neptune and Jupiter is consistent with the fact
that the number of the known Centaur objects at different heliocentric
distances has been found  not  to change substantially with the distance.
Bearing in mind that the observational selection is the larger, the
bigger is the distance of an object both from the Sun and the observer,
the above fact implies that the number of Centaurs should sharply
increase toward the Kuiper belt.

As can be seen from Figs.~4a and 4b, our simulations apparently
do not explain the known comets with the largest 
eccentricities $e>0.9$ or inclinations $i >40^\circ$. Such objects,
which mostly belong to the Saturn,- Uranus,- and Neptune-family comets,
appear to be very rare in the dynamical evolution of
short-period comets. A similar conclusion was made by Levison \& Duncan 
(1997) who argue that
the majority of such comets (of Halley type) could be produced 
by a journey  from the Oort cloud, and not the Kuiper belt.
    
Fig.~5a to 5d show the distributions of the cometary populations governed
by all four giant planets in semimajor axis, distance of pericenter, 
distance of apocenter, and heliocentric distance, respectively.

Fig.~5a demonstrates that the resonant structure in the entire cometary
population is preserved despite the gravitational influence of all four
giant planets. 
The resonant structure is especially rich in the outer part of the
Neptunian cometary belt, where the influence of the other giant planets
is somewhat weakened. On the other hand, the resonant structure
in the distribution  of comets with $a<30$ AU is much more smoothed,
which can be explained by a strong interaction with all giant
planets, including the most massive ones.

Fig.~5b (distribution of simulated comets in pericenter distance)
reveals four major maxima indicating the existence of the four
cometary belts associated with the giant planet orbits.
The separation into four belts becomes more evident for
comets with large $a$ (say, with $39<a<75$ AU, as can be seen in Fig.~4a),
because such comets are dynamically governed by the planet whose orbit 
turns out to be located nearby the comet's pericenter.

The general distribution of comets in distance of apocenter
(Fig.~5c) does not reveal appreciable concentrations to any planet's
orbit, except that of Jupiter.
The other, less contrast peaks are hard to be seen, because
their apocenter maxima are easily destroyed by the influence of the
planets (even the apocenter branch of the Neptunian belt is somewhat
dissolved by the three innermost giant  planets).

Distribution of comets in  heliocentric distance (Fig.~5d) reveals
a density peak near Neptune and another one near Jupiter, i.e.
around the boundaries where those planets are the only hosts. Absence
of noticeable density peaks associated with the orbits of Uranus and Saturn
is not surprising, because those peaks are overlapped by a vast number
of comets belonging to the Neptunian cometary belt. To illustrate this,
we show in Fig.~5d, in one-planet approximation, the distribution of
comets near each giant planet orbit. We assume that the density maxima in 
each belt are proportional to the transfer functions found by OGT (obviously,
we need this assumption only to illustrate how the different belts could be
populated  relative each other).

Our 36 runs performed in the four-planet approximation allow to construct
a steady-state  distribution consisting of $25.7\times 10^6$ positions of
test comets. Of this number of comets, only 815 penetrated into the zone
of visible comets, with distances of pericenter  less than 2.5 AU.
Interestingly, this number turns out not to differ much from the total
number of Jupiter family comets estimated, with accounting
for observational selection, to be $800\pm 300$ (Fern\'andez et al. 1999).
This implies that our simulations indicate the total number
of comets in the Solar system, with the size of several km (typical
for Jupiter family comets), to be as large as 20-30 millions.

According to  our simulations,
the number of gravitationally scattered comets of the Neptunian belt is
as large as $(10-20)\cdot 10^6$. Using recent observations
(Marsden 1999), which indicate that the
number of kuiperoids exceeds the number of scattered Neptunian comets by
a factor of 50, we estimate the total number of kuiperoids to
be $5\cdot 10^8-10^9$ bodies, which is fairly consistent
with $8\cdot 10^8$ inferred by Jewitt (1999) from available observations.
   
\section*{4. Discussion and Conclusions}

One-planet approximation (the Sun plus one planet) employed in Sec.~2 
suggests that each giant planet can host a {\it cometary belt} --
an accumulation of comets associated with the planet's orbit. This is
a non-trivial result: in principle, the distribution
of comets governed by the gravitational fields of the Sun and the planet
could be alike the main asteroidal belt, i.e. not to have any concentration
toward the planet's orbit. The above accumulation has the dynamical nature
implying that each comet in the belt is either gravitationally scattered 
predominantly on this planet or is in a resonance with it. The major 
problem would be to verify whether this accumulation found in one-planet
approximation is survived when the influence of all giant planets is taken 
into account. As shown in Sec.~3, this is indeed the case for a 
cometary population originating in the Kuiper belt and eventually distributed
in the steady-state between the orbits of Neptune and Jupiter.  Although the
cometary belts of all four giant planets can be traced using the
distributions in semimajor axis, distance of pericenter, and heliocentric
distance, the belts are overlapped. The four-planet approximation indicates
that only a tiny fraction of comets is able to penetrate from an outer
planet's zone into the zone of the nearest inner neighbor. As a result,
different cometary  constituents are seen in the superposition of the belts 
with a different confidence:
the Uranian and Saturnian belts are barely seen having as a background
the copious Neptunian belt, and only the latter (plus, in part, the Jovian
belt) appear to be well pronounced. 
Nevertheless, as shown in Sec.~2, the  comets of {\it each} belt, regardless of
its richness, are concentrated to the host's orbit.
Despite the destructive influence
of the four giant planets, the cometary belts maintain their major features
found in one-planet approximation: (i) the belts are the more clear
separated, the larger are semimajor axes of comets, and (ii) the resonant
accumulations and gaps, although lose a little, are still well delineated.

To describe the spatial distribution of comets in the Solar system,
astronomers traditionally use such a term as  `cometary family' 
(e.g. Jupiter family comets). This term, although introduced on a purely
observational basis, turned out to be very valuable as it helped to reveal
the first structures in the cometary population. 
However, since it has been found from numerical simulations that the
cometary population in the zone of giant planets is very populous (Levison
\& Duncan 1997; OGT), it becomes more and more clear that the part of this
population observed in the form of the above cometary families is no more 
than the `tip of the iceberg'. In this paper, we give new evidence in favor
of this. Moreover, the distribution of the cometary populations between 
Jupiter and Neptune simulated in OGT and the present paper is important 
to compute the distribution of dust in the outer Solar system (GOTM 1999). 

     Each cometary family is characterized by the same apocenters of comets 
as the orbit of its host planet.  Therefore, each family of comets turns 
out to be a part of a more general dynamical substance described in this 
paper -- the cometary belt. Such belts, as shown
above, should exist near each giant planet's orbit. The basic features 
of each cometary belt are determined by its
dynamical interaction with the gravitational field of the host planet, and 
these features, as distinct from those in cometary families, do not depend
upon observational selection. The dynamical term  `cometary belt' seems to be
more justifiable and helpful than the observational term `cometary
family'. The latter is meaningful to characterize the visible part of a
cometary belt, which steady grows as soon as more and more faint objects 
are registered with improving techniques. Further 
simulations would be highly desirable to separate possible 
contributions to the cometary belts from the Kuiper belt and Oort cloud.

\vspace{0.1truein}
{\it Acknowledgements}. This work has been supported by NASA Grant NAG5-7065 
to George Mason University. Discussions with, and an invariable support of,
John Mather are highly appreciated. N.G.acknowledges the NRC-NAS
associateship. We are very thankful to Alexander Krivov, the referee, for a
number of helpful suggestions.
 
\newpage

\centerline{\bf References}
\def\ref#1  {\noindent \hangindent=24.0pt \hangafter=1 {#1} \par}
\smallskip
\ref{Fern\'andez, J.A., Ip, W.-H. 1983, Icarus, 54, 377-387}
\ref{Fern\'andez, J.A., Tancredi, G., Rickman, H., Licandro, J. 1999,
``Asteroids, Comets, Meteors" (Abstracts of the International  Meeting at
Cornell Univ., July 26-30), p.~99; Astronomy \& Astrophysics (submitted)}
\ref{Gorkavyi, N.N., Ozernoy, L.M., Taidakova, T., Mather, J.C. ($\equiv$
GOTM) 1999, Four circumsolar dust belts in the outer Solar system associated
with the giant planets. ``Asteroids, Comets, Meteors" (Abstracts of the
International Meeting at Cornell University, July 26-30), p.~131}
\ref{Jewitt, D. 1999, Kuiper belt objects, Ann. Rev. Earth. Planet. Sci., 
27, 287-312}
\ref{Levison, H.F., Duncan M.J. 1997. From the Kuiper belt to 
Jupiter-family comets: the spatial distribution of ecliptic comets.
Icarus 127, 13-32 (LD)}. 
\ref{Malhotra, R., Duncan, M., \& Levison, H. 1999. Dynamics of the Kuiper 
belt objects. In {\it Protostars and Planets IV} 
(in press) $=$ astro-ph/9901155}
\ref{Marsden, B.G. 1999, Minor Planet Center, Ephemerides and Orbital Elements
{\tt http://cfa-www.harvard.edu/iau/Ephemerides/index.html}}
\ref{ Ozernoy, L.M., Gorkavyi, N.N., Taidakova, T. ($\equiv$ OGT) 
2000, Large scale structures in the outer Solar system: 
I.~Cometary belts with resonant features associated with the giant planets.
Icarus (submitted) (OGT) (an earlier version was posted as astro-ph/9812479}
\ref{Taidakova, T. 1997, A new stable method for long-time integration in an
N-body problem. {\it Astronomical Data Analyses,
Software and Systems VI}, ed. G. Hunt \& H.E.Payne, (San Francisco:
 ASP), ASP Conf. Ser. 125, p. 174-177}
\ref{Torbett, M.V. 1989, Chaotic motion in a comet disk beyond Neptune: 
the delivery of short-period comets. Astron. J. 98, 1477-1481}
\newpage
\centerline{\bf Figure Captions}
\vspace{0.1in}
{\bf Figure 1}.

{\bf a}.  2D density of a cometary belt in coordinates `eccentricity --
semimajor axis'. 
To represent the number of comets in each cell, a logarithmic grey 
scale is used, i.e. each shade differs 10-fold from the neighboring one. 
Heavy curves represent the boundaries of the 
crossing zone, and the region above the dashed line
is the zone of visible comets ($q<0.5R_{pl}$). 
Numerous resonant gaps are seen.  

{\bf b}.  2D  density  of  a  cometary belt in coordinates `inclination
angle -- semimajor axis'. The same logarithmic grey 
scale as in {\bf a} is used. Numerous resonant gaps are clearly seen at all
inclinations.  

\vspace{0.1in}
{\bf Figure 2}.

{\bf a}.  Distribution of comets in semimajor axis, with a bin size
$\Delta a=0.005~R_{planet}$.
Various resonant gaps are indicated by arrows. 
The region shown by dashed lines is occupied by
visible comets, whose perihelion distances are the smallest.

{\bf b}. Distribution of comets in the distance of pericenter.
Dashed line indicates the region of visible comets concentrated at the 
left edge of the distribution.

{\bf c}.  Distribution of comets in the distance of apocenter.
Visible comets form a peak near the planet's orbit.

{\bf d}. Surface density of the cometary population as a function of
heliocentric distance. Visible comets are concentrated near
or inside the host planet's orbit.
 
\vspace{0.1in}
{\bf Figure 3}.

 Surface density of cometary populations (logarithmic scale) as a 
function of the heliocentric distance
shown for all four giant planets (in  one-planet approximation).
\newpage
{\bf Figure 4}.

{\bf a}. 2D density of the simulated cometary population of the Solar
system in coordinates `eccentricity -- semimajor axis'
(the four-planet approximation). Four cometary belts 
of the giant planets can be seen.
The boundaries of the crossing  zones are shown by heavy lines, 
and the region occupied by
visible comets ($q<2$ AU) is located above the  dashed line.
Crosses stand for asteroids of
the main belt (the first 100 objects of the list), small triangles stand for 
short-periodic comets (112 objects), large triangles stand for
Centaurs (15 objects), and diamonds stand for the Kuiper belt objects (191).

{\bf b}.  2D  density of the four cometary belts in coordinates `inclination
angle -- semimajor axis'(the four-planet approximation).  
Designations are the same as in {\bf a}.

\vspace{0.1in}
{\bf Figure 5}.

{\bf a}.  Distribution of simulated comets of the Solar system in semimajor 
axis (the four-planet approximation). Arrows indicate various  Neptunian 
resonances. A region inside  Neptune's orbit, where the strong scattering 
zones of different planets are overlapped, looks more uniform than a well 
structured outermost zone.

{\bf b}. Distribution of comets in the distance of pericenter (the four-planet 
approximation). Arrows indicate the four well pronounced peaks
which correspond to the four cometary belts. Dashed line is for comets 
with $39<a<75$ AU.

{\bf c}.  Distribution of comets in the distance of apocenter
(the four-planet approximation).
There is a noticeable peak only around Jupiter's orbit, 
the other peaks in this distribution are associated with
various local resonances and  diffusion accumulations
of comets.

{\bf d}. Surface density of the cometary population as a function of
heliocentric distance (the four-planet approximation).
Curve 1 is the four-planet approximation, and curve 2 is the sum of
4 one-planet approximations (dotted line shows the Neptunian belt, dashed
line shows the Uranian belt, dash-dotted line stands for the Saturnian
belt, and least-populated belt is that of Jupiter).

\end{document}